# Spatially inhomogeneous field induced harmonic radiation in solids

Xiaoxue Zhang[1,2,†], Shiyu Liu[1,2,†] and Chengpu Liu[1,2,*]
[1]State Key Laboratory of Ultra-intense Laser Science and Technology, Shanghai Institute of Optics and Fine Mechanics, Chinese Academy of Sciences, Shanghai 201800, China
[2] Center of Materials Science and Optoelectronics Engineering, University of Chinese Academy of Sciences, Beijing 100049, China

By theoretical derivation, we constructed an inhomogeneous coefficient equation to correctly describing harmonic radiation in solids induced by a spatially inhomogeneous field, where the widely used semiconductor Bloch equation fails. This equation has superiority over the semiconductor Bloch equation with good applicability to both homogeneous and inhomogeneous fields. Using graphene as an example, it is found that under inhomogeneous field driving, even-order harmonics occur with an enhancing tendency as the field inhomogeneity increases. As for the second-order harmonic, its intensity dependence is consistent with the prediction from the perturbation theory, and its wavelength dependence can use to directly distinguish the relative contribution of intraband and interband transitions. The inhomogeneous coefficient equation provides a direct theoretical analysis tool for elucidating the physical mechanism of inhomogeneous field induced harmonic radiation in solids.

## I. INTRODUCTION

High-order harmonic generation (HHG), an important research hotspot in the field of light-matter interaction, not only enables the development of stable and compact coherent light sources on a chip, but also allows for the observation and manipulation of ultrafast electron motions on the sub-femtosecond scale by generating attosecond pulses [1-4]. Compared to gaseous materials, solid materials because of their high density and periodic structure can reduce the requirement for laser intensity and thus increase the conversion efficiency of HHG. Benefiting from the breakthroughs in ultra-intense and ultra-short laser technology, the study of HHG in solid materials [5-9] has experienced a rapid development since it was experimentally observed in ZnO crystal [10]. Initially, the study of HHG was mainly based on spatially homogeneous laser field, while with the intensive study of nanophotonics and surface plasmonics, spatially inhomogeneous laser filed can be formed by different shapes of metallic nanostructures that can generate localized plasmonic enhancement [11]. Under inhomogeneous fields, the HHG exhibits unique nonlinear optical response such as harmonic enhancement and the cutoff frequency increase. Although inhomogeneous field induced harmonic radiation has been experimentally [12-16] explored in both gases and solids, its theoretical analysis has mainly focused on gases [17-27], with a relative lack of theoretical model construction and mechanism elucidation for solids.

Present models for harmonic radiation in solids induced by a spatially inhomogeneous field [28-30] are mainly based on the time-dependent Schrödinger equation (TDSE) or the time-dependent density-functional theory (TDDFT) [31,32] to calculate the total current which is difficult to directly distinguish between intraband and interband harmonics. In contrast, although the conventional semiconductor Bloch equation (SBE) [33] has the ability to distinguish between intraband and interband harmonics, it cannot be applied to inhomogeneous field which breaks the translational symmetry of Bloch's theorem. In this context, starting from the original TDSE, we construct the inhomogeneous coefficient equation (ICE) that can not only be used for describing harmonic radiation in solids under a spatially inhomogeneous field, but also for directly distinguishing the intraband and interband components of harmonics. As an example, we applied ICE to graphene and calculated the harmonic spectrum when an inhomogeneous femtosecond laser irradiating graphene normally. It was found that extra even-order harmonics are generated, and as the field inhomogeneity increases, the even-order harmonics enhance and follow the predictions of the lowest-order perturbation theory. In addition, we specifically investigate the effect of wavelength on the intraband and interband components of the second-order harmonic. As the wavelength increases, the contribution of the intraband transition to the second-order harmonic gradually increases.

The paper is constructed as follows: Section II shows the derivation of the ICE under a spatially inhomogeneous field. In Section III, ICE is solved in graphene and the generation of even-order harmonics is discussed. The conclusion is in Section IV.

## II. THEORY

For a linearly polarized incident laser, since the electron motion is predominantly confined along the direction of the electric field polarization, we derive and compute in the one-dimensional direction. The vectorial nature of the incident field is not considered within the one-dimensional framework, and will be further investigated in the future. According to the finite element simulation results in Ref. [19], we consider only the linear term in the series expansion of the inhomogeneous field function, which serves as a first-order approximation of the near-field around the metal nanoparticle. The expression for the spatially inhomogeneous field is given by [23-29,34,35] $F(x,t) = F(t)(1+\varepsilon x)$, where $F(t)$ is a spatially independent incident laser field (homogeneous field) with polarization along the $x$ direction. $\varepsilon$ is a inhomogeneous parameter with suitable value to keep $\varepsilon x \ll 1$.

†These authors contributed equally to this work.
*Contact author: chpliu@siom.ac.cn

The Hamiltonian under a spatially inhomogeneous field is

$$\hat{H}(t) = \hat{H}_0 - exF(t) - \frac{1}{2}e\varepsilon x^2 F(t). \quad (1)$$

Here $\hat{H}_0$ is the unperturbed Hamiltonian of electrons. The other two terms correspond to the electric dipole and the electric quadrupole, respectively, with the derivation process detailed in the Appendix A. Substituting Eq. (1) into TDSE,

$$i\hbar \frac{\partial}{\partial t} \psi(x,t) = \hat{H}(t)\psi(x,t). \quad (2)$$

The wave function $\psi(x,t)$ can be expanded by the orthogonal normalized Bloch wave function $\phi_{m,k}(x)$ as

$$\psi(x,t) = \sum_m \int_{BZ} a_m(k,t)\phi_{m,k}(x)dk, \quad (3)$$

with the expansion coefficients $a_m(k,t)$. For simplicity, the subscript $x$ in $k_x$ is omitted. $\phi_{m,k}(x) = \frac{1}{\sqrt{N}} u_{m,k}(x)e^{ikx}$, where $u_{m,k}(x)$ is the periodical function with the number of unit cells of crystal $N$, satisfies $\int_{crystal} \phi_{m,k}^*(x)\phi_{m',k'}(x)dx = \delta_{m,m'}\delta_{k,k'}$.

By inserting Eq. (3) into Eq. (2), multiplying by $\phi_{m,k}^*(x)$ and integrating in real space, Eq. (2) can be rewritten as

$$i\hbar \frac{\partial}{\partial t} a_m(k,t) = E_m(k)a_m(k,t)$$
$$-eF(t)\sum_{m'}\int_{BZ} a_{m'}(k',t)\xi_1(m,m',k,k')dk'$$
$$-\frac{1}{2}eF(t)\sum_{m'}\int_{BZ} a_{m'}(k',t)\varepsilon\xi_2(m,m',k,k')dk', \quad (4)$$

with

$$\xi_1(m,m',k,k') = \frac{1}{N}\int_{crystal} u_{m,k}^*(x)xu_{m',k'}(x)e^{i(k'-k)x}dx \quad (5)$$

and

$$\xi_2(m,m',k,k') = \frac{1}{N}\int_{crystal} u_{m,k}^*(x)x^2 u_{m',k'}(x)e^{i(k'-k)x}dx. \quad (6)$$

Here

$$\xi_1(m,m',k,k') = -i\delta_{m,m'}\nabla_k\delta_{k,k'} + \delta_{k,k'}D(m,m',k,k'), \quad (7)$$

which can be obtained from the widely used SBE [33], with

$D(m,m',k,k') = i\int_{cell} u_{m,k}^*(x)e^{i(k'-k)x}\nabla_k u_{m',k'}(x)dx$. It is worth noting that the integral over the crystal is transformed into an integral over the unit cell by using

$$\frac{1}{N}\int_{crystal} u_{m,k}^*(x)e^{i(k'-k)x}\nabla_k u_{m',k'}(x)dx$$
$$= \sum_{n=0}^{N-1}\frac{1}{N}e^{in(k'-k)R}\int_{cell} u_{m,k}^*(x)e^{i(k'-k)x}\nabla_k u_{m',k'}(x)dx$$
$$= \delta_{k,k'}\int_{cell} u_{m,k}^*(x)e^{i(k'-k)x}\nabla_k u_{m',k'}(x)dx. \quad (8)$$

Next, focusing on Eq. (6), we transform to the reciprocal space representation. Because

$$\nabla_k^2\left[u_{m,k}(x)e^{ikx}\right] = 2ixe^{ikx}\nabla_k u_{m,k}(x) + e^{ikx}\nabla_k^2 u_{m,k}(x) - x^2 u_{m,k}(x)e^{ikx}, \quad (9)$$

the first term on the right-hand side of Eq. (9) needs further treatment,

$$2\nabla_k\left[e^{ikx}\nabla_k u_{m,k}(x)\right] = 2ixe^{ikx}\nabla_k u_{m,k}(x) + 2e^{ikx}\nabla_k^2 u_{m,k}(x). \quad (10)$$

Organizing Eqs. (9) and (10), we can get

$$x^2 u_{m,k}(x)e^{ikx} = 2\nabla_k\left[e^{ikx}\nabla_k u_{m,k}(x)\right] - e^{ikx}\nabla_k^2 u_{m,k}(x) - \nabla_k^2\left[u_{m,k}(x)e^{ikx}\right]. \quad (11)$$

By substituting Eq. (11) into Eq. (6),

$$\xi_2(m,m',k,k') = \frac{1}{N}\int_{crystal} u_{m,k}^*(x)e^{-ikx}\left\{2\nabla_{k'}\left[e^{ik'x}\nabla_{k'} u_{m',k'}(x)\right] - e^{ik'x}\nabla_{k'}^2 u_{m',k'}(x) - \nabla_{k'}^2\left[u_{m',k'}(x)e^{ik'x}\right]\right\}dx. \quad (12)$$

For Eq. (12), the order of integral and derivation is exchanged, and we transform to integrating in the unit cell. Thus

$$\xi_2(m,m',k,k') = -\delta_{m,m'}\nabla_k^2\delta_{k,k'} - \delta_{k,k'}C(m,m',k,k') - 2i\nabla_{k'}\left[\delta_{k,k'}D(m,m',k,k')\right], \quad (13)$$

with $C(m,m',k,k') = \int_{cell} u_{m,k}^*(x)e^{i(k'-k)x}\nabla_k^2 u_{m',k'}(x)dx$.

By inserting Eq. (7) and Eq. (13) into Eq. (4), according to the Divergence theorem, and then rounding off the surface integral over the Brillouin zone. Finally, we can get the inhomogeneous coefficient equation (ICE),

$$i\hbar \frac{\partial}{\partial t} a_m(k,t) = E_m(k) a_m(k,t) - ieF(t)\nabla_k a_m(k,t)$$
$$- eF(t)\sum_{m'} a_{m'}(k,t) D(m,m',k)$$
$$+ \frac{1}{2}\varepsilon eF(t)\nabla_k^2 a_m(k,t) \quad (14)$$
$$+ \frac{1}{2}\varepsilon eF(t)\sum_{m'} a_{m'}(k,t) C(m,m',k)$$
$$- i\varepsilon eF(t)\sum_{m'} D(m,m',k)\nabla_k a_{m'}(k,t),$$

which is the basis for the subsequent study of inhomogeneous field-induced harmonic radiation in solids, where $D(m,m',k) = i\int_{cell} u_{m,k}^*(x)\nabla_k u_{m',k}(x)dx$ and $C(m,m',k) = \int_{cell} u_{m,k}^*(x)\nabla_k^2 u_{m',k}(x)dx$. It can be seen that at $\varepsilon = 0$ the last three terms on the right hand side of Eq. (14) vanish and the equation can be reorganized into the form of a density matrix (SBE). However, at $\varepsilon \neq 0$, it is difficult to construct the ICE in the form of a density matrix due to the presence of second order derivatives, hence it is directly solved numerically.

### III. APPLICATION AND DISCUSSION

The spatially homogeneous linearly polarized laser incident normally on the metal nanostructure produces an inhomogeneous field which interacts with graphene, and the ICEs under the two-band model (c-conduction band and v-valence band) are

$$i\hbar \frac{\partial}{\partial t} a_c(k_x,t) = E_c(k_x) a_c(k_x,t) - ieF(t)\nabla_{k_x} a_c(k_x,t) - eF(t)\left[a_c(k_x,t) D_{cc}(k_x) + a_v(k_x,t) D_{cv}(k_x)\right]$$
$$+ \frac{1}{2}\varepsilon eF(t)\nabla_{k_x}^2 a_c(k_x,t) + \frac{1}{2}\varepsilon eF(t)\left[a_c(k_x,t) C_{cc}(k_x) + a_v(k_x,t) C_{cv}(k_x)\right] - i\varepsilon eF(t)\left[D_{cc}(k_x)\nabla_{k_x} a_c(k_x,t) + D_{cv}(k_x)\nabla_{k_x} a_v(k_x,t)\right],$$
$$i\hbar \frac{\partial}{\partial t} a_v(k_x,t) = E_v(k_x) a_v(k_x,t) - ieF(t)\nabla_{k_x} a_v(k_x,t) - eF(t)\left[a_c(k_x,t) D_{vc}(k_x) + a_v(k_x,t) D_{vv}(k_x)\right]$$
$$+ \frac{1}{2}\varepsilon eF(t)\nabla_{k_x}^2 a_v(k_x,t) + \frac{1}{2}\varepsilon eF(t)\left[a_c(k_x,t) C_{vc}(k_x) + a_v(k_x,t) C_{vv}(k_x)\right] - i\varepsilon eF(t)\left[D_{vc}(k_x)\nabla_{k_x} a_c(k_x,t) + D_{vv}(k_x)\nabla_{k_x} a_v(k_x,t)\right].$$
(15)

The Hamiltonian of graphene [36,37] in the tight-binding model is

$$\hat{H}_0 = \begin{bmatrix} 0 & -\gamma_0 f(\mathbf{k}) \\ -\gamma_0 f^*(\mathbf{k}) & 0 \end{bmatrix}. \quad (16)$$

$\gamma_0 = 3\text{eV}$ is the nearest neighbor hopping parameter, $f(\mathbf{k}) = \exp(i\frac{ak_x}{\sqrt{3}}) + 2\exp(-i\frac{ak_x}{2\sqrt{3}})\cos(\frac{ak_y}{2})$, with lattice constant $a = 0.246\text{nm}$. As the incident laser is polarized along the Γ-M direction, $k_y = 0$. Due to the existence of the singularity at the Dirac point in the Γ-K direction and the divergence of the transition dipole moment, further study on this direction will be considered in the future. By diagonalizing Eq. (16), one obtains the conduction band energy $E_c(k_x) = \gamma_0 |f(k_x,k_y=0)|$ and the valence band energy $E_v(k_x) = -\gamma_0 |f(k_x,k_y=0)|$. The corresponding periodical functions are

$$u_{c,k_x}(x) = \frac{1}{\sqrt{2}}\begin{pmatrix} -\exp(i\theta_{k_x}/2) \\ \exp(-i\theta_{k_x}/2) \end{pmatrix}, u_{v,k_x}(x) = \frac{1}{\sqrt{2}}\begin{pmatrix} \exp(i\theta_{k_x}/2) \\ \exp(-i\theta_{k_x}/2) \end{pmatrix},$$
(17)

with the phase angle $\theta_{k_x} = \arg(f(k_x,k_y=0))$. Through Eq. (17), we can obtain

$$D_{cc}(k_x) = D_{vv}(k_x) = 0, D_{cv}(k_x) = D_{vc}(k_x) = \frac{1}{2}\nabla_{k_x}\theta_{k_x},$$
$$C_{cc}(k_x) = C_{vv}(k_x) = -\frac{1}{4}\left(\nabla_{k_x}\theta_{k_x}\right)^2, C_{cv}(k_x) = C_{vc}(k_x) = -\frac{i}{2}\nabla_{k_x}^2\theta_{k_x}.$$
(18)

The explicit expansions are as below

$$D_{cv}(k_x,k_y=0) = \frac{a}{2\sqrt{3}}\frac{\cos(\sqrt{3}ak_x/2)\cos(ak_y/2) - \cos(ak_y)}{3 + 4\cos(\sqrt{3}ak_x/2)\cos(ak_y/2) + 2\cos(ak_y)},$$
$$C_{cv}(k_x,k_y=0) = \frac{i}{2}\frac{3a^2\left[1 + 2\cos(ak_x/\sqrt{3})\right]\cos(ak_y/2)\left[1 + 2\cos(ak_y)\right]\sin(ak_x/2\sqrt{3})}{2\left[3 + 4\cos(\sqrt{3}ak_x/2)\cos(ak_y/2) + 2\cos(ak_y)\right]^2}.$$
(19)

As the inhomogeneity of the spatial field has been translated to the ICEs, the expression for the homogeneous field is $F(t) = F_0 \exp(-2\ln 2 \frac{t^2}{t_p^2})\cos(\omega t)$. $F_0$ is the peak amplitude and peak intensity $I = \frac{1}{2}c_0\varepsilon_0 F_0^2$, with light speed $c_0$, vacuum permittivity $\varepsilon_0$, full width at half maximum (FWHM) $t_p$ and angular frequency $\omega$. In solving Eq. (15), the fourth-order Runge-Kutta method is used for the time-difference term and the fourth-order center difference is applied for the space-difference term, and periodic boundary condition is set. The time step of 0.0005 fs and the number of spatial grid points of 10001 were used in the numerical simulation.

Introducing the interband polarization $\rho_{cv}(k_x,t) = a_c(k_x,t)a_v^*(k_x,t)$ and intraband population of electron $\rho_{c(v)}(k_x,t) = a_{c(v)}(k_x,t)a_{c(v)}^*(k_x,t)$, we can get the intraband current

$$J_{intra}(t) = \frac{-e}{\pi}\int [\rho_c(k_x,t)v_c(k_x) + \rho_v(k_x,t)v_v(k_x)]dk_x, \quad (20)$$

with the group velocity $v_{c(v)}(k_x) = \nabla_{k_x} E_{c(v)}(k_x)/\hbar$, and the interband current

$$J_{inter}(t) = \frac{-e}{\pi}\frac{\partial}{\partial t}\int \rho_{cv}(k_x,t)D_{cv}(k_x)dk_x + \text{c.c.} \quad (21)$$

Compared to the total current calculated by TDSE, the ICE can directly obtain the intraband and interband current components. We multiply the current by a super-Gaussian function to smooth the edge truncation. The harmonic spectrum is proportional to the modulus square of the Fourier transform to transient current $I(\omega) \propto |J_{intra}(\omega) + J_{inter}(\omega)|^2$.

It has been shown that the generation of even harmonics relies on symmetry breaking of the system, which is mainly realized by asymmetric laser field modulation [38,39] or symmetry breaking materials [6,8,40]. For the latter, the generation mechanism mainly originates from the complex transition dipole phase [41]. In order to clearly demonstrate the effect of the spatially symmetry-broken laser field on the harmonic radiation, graphene was chosen for the study. The harmonic spectra of graphene calculated with the ICE under homogeneous and inhomogeneous fields are shown in Fig. 1. In the homogeneous field Fig. 1(a), $\varepsilon = 0$, the results obtained by ICE are consistent with those of SBE, clearly showing only odd-order harmonics generation. When the inhomogeneous parameter $\varepsilon = 0.01/m$ is introduced, as shown in Fig. 1(b), even-order harmonics are generated in addition to odd-order harmonics. Because the introduced inhomogeneous part is a perturbation with respect to the homogeneous part, the resulting even-order harmonics are of low intensity. Increasing the inhomogeneous parameter as $\varepsilon = 0.1/m$, the intensity of even-order harmonics increases.

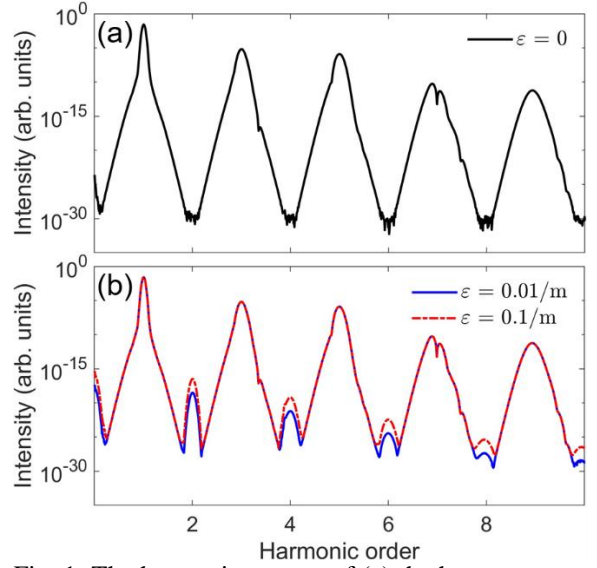

Fig. 1. The harmonic spectra of (a) the homogeneous field and (b) the inhomogeneous field. Laser parameters are 800 nm in wavelength, 25 fs in FWHM, and $2\times10^{11}$ W/cm$^2$ in peak intensity.

Considering the low intensity of the even-order harmonics, our subsequent analysis of the even-order harmonics focuses on the second-order harmonic. The radiation intensity dependence on laser peak intensity for the second-order, third-order and fifth-order harmonics is demonstrated in Fig. 2. From Fig. 2(a), it can be observed that the second-order harmonic intensity increases as the peak intensity increases, and the trend is consistent with perturbation theory, i.e., it shows a dependence on $I^2$. In contrast, in Figs. 2(b) and 2(c), the effect of the inhomogeneous field on the odd-order harmonics is almost negligible. As for the intensity dependence of the third-order harmonic, it is consistent with perturbation theory at smaller peak intensities, but the trend significantly deviates from the $I^3$ dependence after increasing the peak intensity, resulting in a non-perturbative situation. While in the fifth-order harmonic, the larger range of peak intensities is consistent with perturbation theory, after which a saturating behavior is shown, and this feature is in general agreement with the result reported by Yoshikawa et al. [7].

In particular, in order to show that the ICEs are able to directly distinguish between intraband and interband harmonics, we investigate the wavelength dependence of the second-order harmonic, as shown in Fig. 3. When the wavelength is short as shown in Fig. 3(a), the photon energy is high, and the electrons are easy to transition from the valence band to the conduction band, which means that the interband transition probability is strong. Therefore, the contribution of the interband harmonic component to the second-order harmonic generation is larger, compared with the intraband harmonic component. When the wavelength continues to increase, the results are shown in Figs. 3(b) and 3(c). The photon energy decreases, and at the same laser intensity, more photon numbers are needed to make the electron transition, i.e., the transition probability decreases, and the contribution of the interband harmonic component decreases. Furthermore, the increase in wavelength corresponds to the increase in the optical cycle $T$, which means that the electrons have more time to travel within the band, and it is able to accumulate more energy. Thus the contribution of the intraband harmonic component to the second-order harmonic generation is increased. Consequently, for second-order harmonic generation, interband harmonic component dominates at shorter wavelengths and intraband harmonic component dominates at longer wavelengths.

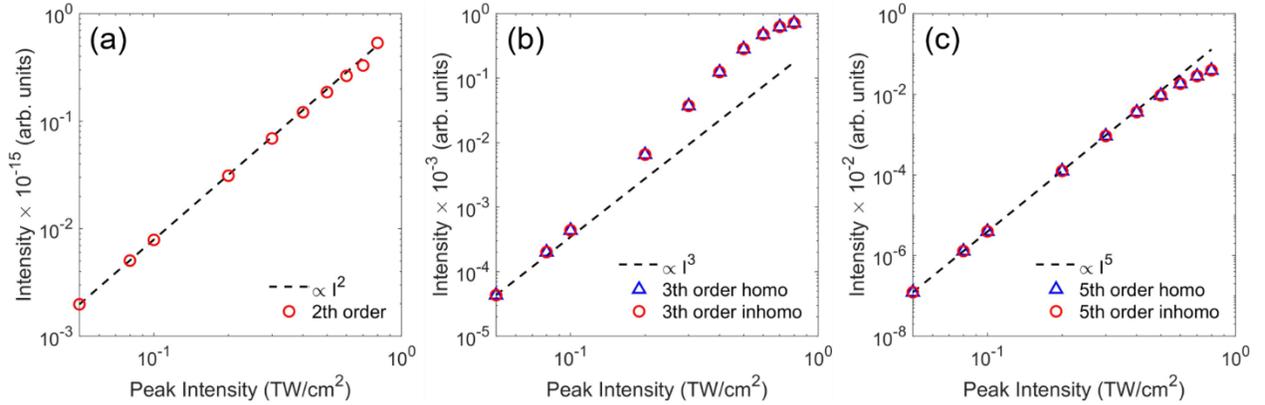

Fig. 2. Log-log plots of the dependence of (a) the second-order, (b) the third-order, and (c) the fifth-order harmonic radiation intensity on the laser peak intensity. The dashed line indicates the perturbation line of the corresponding order. The red circle is under the inhomogeneous field $\varepsilon = 0.1/\mathrm{m}$, and the blue triangle is under the homogeneous field $\varepsilon = 0$. Except for the laser intensity, the other parameters remain fixed.

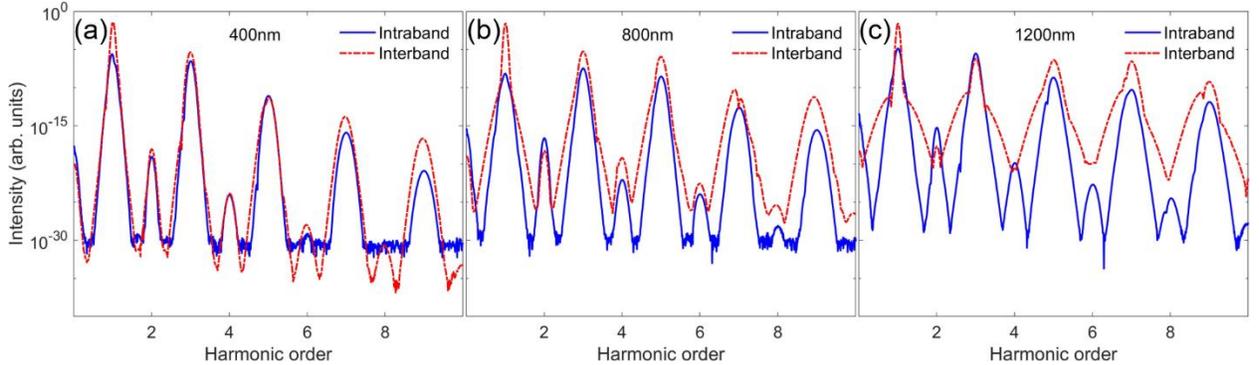

Fig. 3. The intraband and interband harmonic spectra at wavelengths of (a) 400 nm, (b) 800 nm, and (c) 1200 nm. The inhomogeneous parameter $\varepsilon = 0.1/\mathrm{m}$ and the laser intensity $2 \times 10^{11}$ W/cm$^2$ are fixed, and the multiple of the FWHM with respect to the optical cycle $T = \dfrac{2\pi}{\omega}$ is fixed.

## IV. CONCLUSIONS

In conclusion, we construct the inhomogeneous coefficient equation in order to theoretically investigate harmonic radiation in solids under spatially inhomogeneous fields, which can directly distinguish the intraband and interband components of harmonics. This equation is universally applicable, whether to homogeneous or inhomogeneous fields. As an example of harmonic radiation in graphene, it is demonstrated that extra even-order harmonics can be generated under an inhomogeneous field, as well as that the intensity of even-order harmonics increases with increasing field inhomogeneity. In addition, the second-order harmonic intensity exhibits a perturbative dependence on laser peak intensity, while increasing the incident wavelength results in the gradual predominance of its intraband component. The present study is based on the simplest two-band model, but the inhomogeneous coefficient equation constructed is of adaptability, which can be extended to multi-band systems. The inhomogeneous coefficient equation provides an important theoretical basis for accurately describing the physical mechanism of harmonic radiation in solids in inhomogeneous field, as well as an important reference for future experimental studies.

**ACKNOWLEDGMENTS**

This work is supported by the National Natural Science Foundation of China (12074398, 11674342, 11374318).

**APPENDIX A**

In this Appendix we derive the Hamiltonian under a spatially dependent external field from the minimal-coupling Hamiltonian.

The expression for the spatially dependent external field is $\mathbf{F}(\mathbf{r},t) = F(t)(1+\boldsymbol{\varepsilon}\cdot\mathbf{r})\mathbf{q}$, with $F(t) = \text{Re}\left[F_0 f(t)e^{i\omega t}\right]$. $F_0$ is the field peak amplitude and $f(t)$ is the slowly varying envelope function. Considering only the one-dimensional $x$-direction, the perturbation parameter $\boldsymbol{\varepsilon} = (\varepsilon, 0, 0)$ and the polarization vector $\mathbf{q} = (1, 0, 0)$. Since $\mathbf{F}(\mathbf{r},t) = -\frac{\partial}{\partial t}\mathbf{A}(\mathbf{r},t)$, ignoring higher-order terms of $f(t)$, we can obtain the vector potential as

$$\mathbf{A}(\mathbf{r},t) = \frac{i}{\omega}F(t)(1+\boldsymbol{\varepsilon}\cdot\mathbf{r})\mathbf{q}. \quad (A1)$$

The minimal-coupling Hamiltonian is

$$\begin{aligned}H &= \frac{1}{2m}\left[\mathbf{p} - e\mathbf{A}(\mathbf{r},t)\right]^2 + e\phi(\mathbf{r},t) + V(\mathbf{r})\\ &= H_0 - \frac{e}{2m}\left[\mathbf{p}\cdot\mathbf{A}(\mathbf{r},t) + \mathbf{A}(\mathbf{r},t)\cdot\mathbf{p}\right] + \frac{e^2}{2m}\mathbf{A}^2(\mathbf{r},t) + e\phi(\mathbf{r},t)\\ &\approx H_0 - \frac{e}{2m}\left[\mathbf{p}\cdot\mathbf{A}(\mathbf{r},t) + \mathbf{A}(\mathbf{r},t)\cdot\mathbf{p}\right],\end{aligned} \quad (A2)$$

where $\mathbf{p}$ is the momentum operator, the scalar potential $\phi(\mathbf{r},t) = 0$ and the $\mathbf{A}^2(\mathbf{r},t)$ term is usually small that can be ignored. Electrons transition from the initial state $|\phi_i\rangle$ to the final state $|\phi_f\rangle$, $i \neq f$, hence

$$\langle\phi_f|H|\phi_i\rangle = \langle\phi_f|H_0|\phi_i\rangle - \frac{e}{2m}\langle\phi_f|\mathbf{p}\cdot\mathbf{A}(\mathbf{r},t) + \mathbf{A}(\mathbf{r},t)\cdot\mathbf{p}|\phi_i\rangle. \quad (A3)$$

Because

$$\langle\phi_f|\mathbf{p}\cdot\mathbf{A}(\mathbf{r},t)|\phi_i\rangle = \langle\phi_f|\mathbf{A}(\mathbf{r},t)\cdot\mathbf{p}|\phi_i\rangle - i\hbar\nabla\cdot\mathbf{A}(\mathbf{r},t)\langle\phi_f|\phi_i\rangle, \quad (A4)$$

the second term on the right-hand side of Eq. (A3) can be

$$-\frac{e}{m}\langle\phi_f|\mathbf{A}(\mathbf{r},t)\cdot\mathbf{p}|\phi_i\rangle + \frac{i\hbar e}{2m}\nabla\cdot\mathbf{A}(\mathbf{r},t)\langle\phi_f|\phi_i\rangle. \quad (A5)$$

Substitute Eq. (A1) into Eq. (A5),

$$\frac{i\hbar e}{2m}\nabla\cdot\mathbf{A}(\mathbf{r},t)\langle\phi_f|\phi_i\rangle = -\frac{\hbar e\varepsilon}{2m\omega}F(t)\delta_{fi} = 0, \quad (A6)$$

this term does not contribute to the transition.

Now considering the term of $\mathbf{A}(\mathbf{r},t)\cdot\mathbf{p}$,

$$\begin{aligned}&-\frac{e}{m}\langle\phi_f|\mathbf{A}(\mathbf{r},t)\cdot\mathbf{p}|\phi_i\rangle\\ &= -\frac{ieF(t)}{m\omega}\langle\phi_f|(1+\boldsymbol{\varepsilon}\cdot\mathbf{r})\mathbf{q}\cdot\mathbf{p}|\phi_i\rangle\\ &= -\frac{ieF(t)}{m\omega}\langle\phi_f|\mathbf{q}\cdot\mathbf{p}|\phi_i\rangle - \frac{ieF(t)}{m\omega}\langle\phi_f|(\boldsymbol{\varepsilon}\cdot\mathbf{r})(\mathbf{q}\cdot\mathbf{p})|\phi_i\rangle.\end{aligned} \quad (A7)$$

The first term on the right-hand side of Eq. (A7) corresponds to the electric dipole, because

$$\begin{aligned}-\frac{ieF(t)}{m\omega}\langle\phi_f|\mathbf{q}\cdot\mathbf{p}|\phi_i\rangle &= -\frac{ieF(t)}{m\omega}\langle\phi_f|p_x|\phi_i\rangle\\ &= -\frac{ieF(t)}{m\omega}\langle\phi_f|\frac{m}{i\hbar}[x,H_0]|\phi_i\rangle\\ &= -\frac{eF(t)(E_i - E_f)}{\hbar\omega}\langle\phi_f|x|\phi_i\rangle.\end{aligned} \quad (A8)$$

Due to the energy conservation condition, $\omega = (E_i - E_f)/\hbar$, Eq. (A8) is $-eF(t)\langle\phi_f|x|\phi_i\rangle$. To derive the second term on the right-hand side of Eq. (A7), we decompose $(\boldsymbol{\varepsilon}\cdot\mathbf{r})(\mathbf{q}\cdot\mathbf{p})$ into symmetric and antisymmetric parts with respect to the exchange of $\mathbf{r}$ and $\mathbf{p}$.

$$\begin{aligned}(\boldsymbol{\varepsilon}\cdot\mathbf{r})(\mathbf{q}\cdot\mathbf{p}) &= \frac{1}{2}[(\boldsymbol{\varepsilon}\cdot\mathbf{r})(\mathbf{q}\cdot\mathbf{p}) + (\boldsymbol{\varepsilon}\cdot\mathbf{p})(\mathbf{q}\cdot\mathbf{r})]\\ &+ \frac{1}{2}[(\boldsymbol{\varepsilon}\cdot\mathbf{r})(\mathbf{q}\cdot\mathbf{p}) - (\boldsymbol{\varepsilon}\cdot\mathbf{p})(\mathbf{q}\cdot\mathbf{r})].\end{aligned} \quad (A9)$$

First, we derive the symmetric part corresponding to the electric quadrupole.

$$-\frac{ieF(t)}{2m\omega}\langle\phi_f|(\boldsymbol{\varepsilon}\cdot\mathbf{r})(\mathbf{q}\cdot\mathbf{p})+(\boldsymbol{\varepsilon}\cdot\mathbf{p})(\mathbf{q}\cdot\mathbf{r})|\phi_i\rangle$$

$$=-\frac{ieF(t)}{2m\omega}\sum_{jk}\varepsilon_j q_k\langle\phi_f|r_j p_k + p_j r_k|\phi_i\rangle$$

$$=-\frac{eF(t)}{2\hbar\omega}\sum_{jk}\varepsilon_j q_k\langle\phi_f|r_j[r_k,H_0]+[r_j,H_0]r_k|\phi_i\rangle$$

$$=-\frac{eF(t)}{2\hbar\omega}\sum_{jk}\varepsilon_j q_k\langle\phi_f|[r_j r_k,H_0]|\phi_i\rangle$$

$$=-\frac{eF(t)}{2}\sum_{jk}\varepsilon_j q_k\langle\phi_f|r_j r_k|\phi_i\rangle$$

$$=-\frac{e\varepsilon F(t)}{2}\langle\phi_f|x^2|\phi_i\rangle.$$

(A10)

Then the antisymmetric part corresponds to the magnetic dipole.

$$-\frac{ieF(t)}{2m\omega}\langle\phi_f|(\boldsymbol{\varepsilon}\cdot\mathbf{r})(\mathbf{q}\cdot\mathbf{p})-(\boldsymbol{\varepsilon}\cdot\mathbf{p})(\mathbf{q}\cdot\mathbf{r})|\phi_i\rangle$$

$$=-\frac{ieF(t)}{2m\omega}\langle\phi_f|(\boldsymbol{\varepsilon}\times\mathbf{q})(\mathbf{r}\times\mathbf{p})|\phi_i\rangle \quad (A11)$$

$$=-\frac{ieF(t)}{2m\omega}\langle\phi_f|(\boldsymbol{\varepsilon}\times\mathbf{q})\mathbf{L}|\phi_i\rangle$$

$$=0,$$

as under the one-dimensional $x$-direction, $\boldsymbol{\varepsilon}\times\mathbf{q}=0$.

Finally, by reorganizing, Eq. (A3) can be replaced as

$$\langle\phi_f|H|\phi_i\rangle=\langle\phi_f|H_0-exF(t)-\frac{1}{2}e\varepsilon x^2 F(t)|\phi_i\rangle.$$

(A12)

Thus the Hamiltonian under a spatially dependent external field is $H=H_0-exF(t)-\frac{1}{2}e\varepsilon x^2 F(t)$, including both the electric dipole and the electric quadrupole.